\documentclass[twocolumn,superscriptaddress,showpacs,pra]{revtex4}
\usepackage{epsfig}
\usepackage{color}
\usepackage{delarray}
\usepackage{amsmath, amssymb}
\usepackage{bm}
\usepackage{float}

\begin{document}
\title{Compressive Split-Step Fourier Method\footnote{This article can be cited as: Bayindir, C., 2015, ``Compressive Split-Step Fourier Method", TWMS Journal of Applied and Engineering Mathematics, vol. 5, no. 2, 298-306.}}

\author{Cihan Bay\i nd\i r}
\email{cihan.bayindir@isikun.edu.tr}
\affiliation{Department of Civil Engineering, I\c s\i k University,  \.{I}stanbul, Turkey}

%\date{\today}
\begin{abstract}

In this paper an approach for decreasing the computational effort required for the split-step Fourier method (SSFM) is introduced. It is shown that using the sparsity property of the simulated signals, the compressive sampling algorithm can be used as a very efficient tool for the split-step spectral simulations of various phenomena which can be modeled by using differential equations. The proposed method depends on the idea of using a smaller number of spectral components compared to the classical split-step Fourier method with a high number of components. After performing the time integration with a smaller number of spectral components and using the compressive sampling technique with $l_1$ minimization, it is shown that the sparse signal can be reconstructed with a significantly better efficiency compared to the classical split-step Fourier method. Proposed method can be named as compressive split-step Fourier method (CSSFM). For testing of the proposed method the Nonlinear Schr\"{o}dinger Equation and its one-soliton and two-soliton solutions are considered.

\end{abstract}
\maketitle

%%%%%%%%%%%%%%%%%%%%%%%%%%%%%%% main %%%%%%%%%%%%%%%%%%%%%%%%%%%%%
%\begin{section}{Introduction}
\section{Introduction}
Spectral methods is one of the very widely used class of numerical solutions in computational mathematics. Some examples can be seen in \cite{bay2009, canuto,Karjadi2010,Karjadi2012,trefethen}. While spatial derivatives are calculated using spectral techniques, time integration is performed using schemes such as Adams-Bashforth and Runge-Kutta etc \cite{bay_sun3, bay_sun2, demiray}. One of the very efficient methods is the split-step Fourier method (SSFM) which was originally proposed in \cite{hardin}. In SSFM the time integration is performed by time stepping of the exponential function for an equation which includes a first order time derivative. SSFM is widely used in many branches of applied sciences including but not limited to optics, acoustics, oceanography.

It is known that majority of the signals in nature and engineering devices are sparse. Therefore the compressive sampling technique can be thought as a very efficient tool for measuring or simulating the such signals. In this paper it is shown that the efficiency of the compressive sampling technique can also be used for the improvement of the SSFM. For this purpose Nonlinear Schr\"{o}dinger equation and its one- and two-soliton solutions are considered.

The method proposed in this paper depends on the idea of using a smaller number of spectral components in SSFM. Using a smaller number of spectral components and using the compressive sampling technique, it shown that the numerical integration can be performed with a significantly better efficiency compared to the classical SSFM. The proposed method is named as the compressive split-step Fourier method (CSSFM). The sparsity property of the solitons in the time domain is used and ${l_1}$ minimization technique of the compressive sampling algorithm is utilized.  

Starting from the initial conditions time integration is performed with SSFM and CSSFM. It is shown by using the CSSFM, time integration can be performed with a significantly better efficiency compared to the SSFM with a high number of spectral components. Also it is shown that the accuracy difference between two models is of negligible importance. Therefore it is shown that the proposed CSSFM can be a very efficient tool in computational mathematics.
\section{Methodology}

\subsection{Review of the Nonlinear Schr\"{o}dinger Equation}
%\begin{doublespace}
Nonlinear Schr\"{o}dinger equation (NSE) can be written as
\begin{equation}
i\eta_t = \eta_{xx} + 2 \left|\eta \right|^2 \eta
\label{eq01}
\end{equation}
where $x,t$ is the spatial and temporal variables, $i$ denotes the imaginary number and $\eta$ is complex amplitude. NSE is widely used in applied sciences and engineering to describe various phenomena including but not limited to weakly nonlinear ocean waves \cite{zakharov}, pulse propagation in optical fibers and quantum state of a physical system. Integrability of the NSE is studied extensively within last forty years and some exact solutions of the NSE is derived. In this paper one-soliton and two-soliton solutions of the NSE are considered. One-soliton solution of the NSE can be written as \cite{bogomolov,taha}
\begin{equation}
\begin{split}
\eta(x,t)=2A & \exp{\{-i[2x-4\left(1-A^2\right)t+\pi/2] \} }  \\
& .sech(2Ax-8At)
\label{eq02}
\end{split}
\end{equation}
where $A$ is a constant which denotes the amplitude of the NSE and selected as $0.5$ for this study following \cite{taha}.
\newline
Two soliton-solution can be written as \cite{bogomolov, hirota, taha}
\begin{equation}
\eta(x,t)=G(x,t)/F(x,t)
\label{eq03}
\end{equation}
where
\begin{equation}
\begin{split}
F(x,t)=1 &+  a(1,1^*)\exp(\zeta_1+\zeta_1^*)  \\
& +a(1,2^*)\exp(\zeta_1+\zeta_2^*) \\
& +a(2,1^*)\exp(\zeta_2+\zeta_1^*)  \\
& +a(2,2^*)\exp(\zeta_2+\zeta_2^*) \\
& +a(1,2,1^*,2^*)\exp(\zeta_1+\zeta_2+\zeta_1^* + \zeta_2^*)
\end{split}
\label{eq04}
\end{equation}
and
\begin{equation}
\begin{split}
G(x,t)=\exp(\zeta_1) & + \exp(\zeta_2)  + a(1,2,1^*)\exp(\zeta_1+\zeta_2+\zeta_1^*) \\
& + a(1,2,2^*)\exp(\zeta_1+\zeta_2+\zeta_2^*)
\label{eq05}
\end{split}
\end{equation}
where $^*$ denotes the complex conjugate. The parameters in these equations are defined as \cite{bogomolov, taha}
\begin{equation}
\begin{split}
& a(i,j^*)= (P_i + P_j^* )^{-2},\\
& a(i,j)= (P_i - P_j )^{2}, \\
& a(i^*,j^*)= (P_i^* - P_j^* )^{2} \\
& a(i,j,k^*)= a(i,j)a(i,k^*)a(j,k^*), \\
& a(i,j,k^*,l^*)= a(i,j)a(i,k^*)a(i,l^*)a(j,k^*)a(j,l^*)a(k^*,l^*).
\end{split}
\label{eq06}
\end{equation}
and
\begin{equation}
\zeta_j=P_jx-\Omega_jt-\zeta_j^{(0)}, \ \Omega_j=iP_j^2.
\label{eq07}
\end{equation}
where $i$ in the last expression is the imaginary number. The constant parameters in these expressions are taken as
\begin{equation}
P_1=4-2i, \  \ P_2=3+i, \  \ \zeta_1^{(0)}=-9.04,  \  \ \zeta_2^{(0)}=2.1.
\label{eq08}
\end{equation}
as in \cite{bogomolov, taha}. Also the $x$ interval is chosen as L=[-20,20] for simulations following \cite{taha}.
\subsection{Review of the Split-Step Fourier Method} 
SSFM is based on the idea of splitting the equation into two parts, the nonlinear and the linear part \cite{bay2015}. The original form of SSFM can be seen in \cite{hardin}. Since then, researchers have proposed many different versions of the SSFM. A literature view for the proposed modifications to SSFM can be seen in \cite{bogomolov}. For the NSE, the advance in time due to nonlinear part can be written as
\begin{equation}
i\eta_t=2 \left| \eta \right|^2\eta
\label{eq09}
\end{equation}
which can be exactly solved as
\begin{equation}
\tilde{\eta}(x,t_0+\Delta t)=e^{-2i\left| \eta(x,t_0)\right|^2\Delta t}\ \eta(x,t_0)
\label{eq10}
\end{equation}
where $\Delta t$ is the time step. The linear part of the NSE can be written as
\begin{equation}
i\eta_t=\eta_{xx}
\label{eq11}
\end{equation}
Using the Fourier series it is possible to write that \cite{bogomolov, canuto, demiray}
 \begin{equation}
\eta(x,t_0+\Delta t)=F^{-1} \left[e^{ik^2\Delta t}F[\tilde{\eta}(x,t_0+\Delta t) ] \right]
\label{eq12}
\end{equation}
Therefore combining (\ref{eq10}) and (\ref{eq12}), the complete form of the SSFM can be written as
 \begin{equation}
\eta(x,t_0+\Delta t)=F^{-1} \left[e^{ik^2\Delta t}F[ e^{-2i\left| \eta(x,t_0) \right|^2\Delta t}\ \eta(x,t_0) ] \right]
\label{eq12}
\end{equation}
Starting from the initial conditions this expression can be solved explicitly. This operation requires two FFTs per time step. 
%\end{doublespace}

\subsection{Review of the Compressive Sampling}
%\begin{doublespace}
\noindent Since its first appearance in literature, compressive sampling (CS) has drawn the attention of many researchers. Today it is widely used in various branches of engineering, applied physics mathematics. Some studies in digital systems such as the development of a single pixel video camera systems or analog-to-digital converter efficiently use the CS algorithm. In this section a brief summary of the CS is sketched.

Let $\eta$ be a $K$-sparse signal with $N$ elements, that is only $K$ out of $N$ elements of the signal are nonzero. Using orthonormal basis functions with transformation matrix ${\bf \Psi}$, $\eta$ can be represented in terms of basis functions. Typical orthogonal transformation used in the literature are the Fourier, wavelet or discrete cosine transforms just to mention few. Therefore it is possible to write $\eta= {\bf \Psi} \widehat{ \eta}$ where $\widehat{ \eta}$ is the coefficient vector. Discarding the zero coefficients of $\eta$, one can obtain $\eta_s= {\bf \Psi}\widehat{ \eta}_s$  where $\eta_s$ is the signal with non-zero components only.

CS algorithm states that a $K$-sparse signal $\eta$ of length $N$ can exactly be reconstructed from $M \geq C \mu^2 ({\bf \Phi},{\bf \Psi}) K \textnormal{ log (N)}$ measurements with a very high probability, where $C$ is a positive constant and $\mu^2 (\Phi,\Psi)$ is the mutual coherence between the sensing basis ${\bf \Phi}$ and transform basis ${\bf \Psi}$ \cite{candes}.

Taking $M$ random projections and using the sensing matrix ${\bf \Phi}$ one can obtain $g={\bf \Phi} \eta$. Therefore the problem can be re-formulated as
\begin{equation}
\textnormal{ min} \left\| \widehat{ \eta} \right\|_{l_1}   \ \ \ \  \textnormal{under constraint}  \ \ \ \ g={\bf \Phi} {\bf \Psi} \widehat{ \eta}
\label{eq14}
\end{equation}
where $\left\| \widehat{ \eta} \right\|_{l_1}=\sum_i \left| \widehat{ \eta}_i\right|$. So that among all signal which satisfies the given constraints, the ${l_1}$ solution of the CS problem is given as  $\eta_{{}_{CS}} ={\bf \Psi} \widehat{ \eta}$. 

$l_1 $ minimization is only one of the alternatives which can be used for this optimization problem. The sparse solutions can also be recovered with the help of other optimization techniques such as re-weighted $l_1 $ minimization or greedy pursuit algorithms \cite{candes}. Details of the CS can be seen in \cite{candes}.  

%\end{doublespace}
\subsection{Proposed Method}
%\begin{doublespace}
%In this study a methodology which can significantly reduce the computational effort required for the spectral simulations of the sparse water waves is offered. 
\noindent In a classical SSFM let $N$ be the number of the spectral components used for representation of a signal. By using $M$ spectral components with $M << N$ and using the CS technique to construct the $N$-component signal from $M$ components, it is possible to obtain a very efficient computational method especially for very long time evolutions. This method can be named as compressive split-step Fourier (CSSFM) method and recently introduced in \cite{bay_sun1, bay_sun2}. The selection of the number $M$ has to be done carefully depending on width of the $K$-sparse wave profile since $M$ need to satisfy the $M=O(K \log(N/K))$ condition of the CS algorithm. Starting from the initial conditions only $M$ spectral components are time stepped. After the time stepping, the $N$ point signal is reconstructed from $M$ components by using the $l_1$ minimization technique of the CS theory. It is shown that the method offered in here can reduce the computational effort significantly compared to the SSFM with $N$ components while the accuracy difference in the results is of negligible importance.

\section{Results and Discussion}
\subsection{Evolution of One-Soliton}
%\begin{doublespace}

\noindent In the Figure~\ref{fig1} below, the $N=1024$ component SSFM and the $M=128$ component CSSFM are compared for one-soliton solution. The initial condition for this simulation is defined by (\ref{eq02}).  The two methods are in excellent agreement as it can be seen in the figure. The normalized root-mean-square difference between two profiles is $0.0025$ for this simulation.

\begin{figure}[h]
\begin{center}
   \includegraphics[width=3.4in]{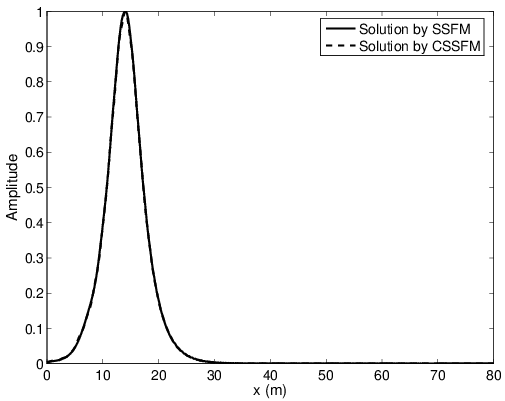}
  \end{center}
\caption{\small Comparison of SSFM with $N=1024$ and  CSSFM with $M=128$ components for one-soliton evolution.}
  \label{fig1}
\end{figure}

In the Figure~\ref{fig2} below, the  $N=1024$ component SSFM and the $M=256$ component CSSFM are compared for one-soliton solution. The two methods are in excellent agreement as it can be seen in the figure. The normalized root-mean-square difference between two profiles is $0.0012$ for this simulation.

\begin{figure}[h]
\begin{center}
   \includegraphics[width=3.4in]{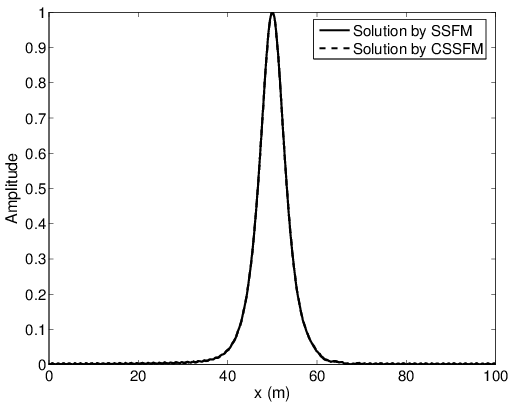}
  \end{center}
\caption{\small Comparison of SSFM with $N=1024$ and  CSSFM with $M=256$ components for one-soliton evolution.}
  \label{fig2}
\end{figure}

All of the results presented in the figures above for one-soliton solution show promising evidence for the accuracy of the proposed method. Additionally the computational effort required to run the various cases are summarized in the Table~\ref{tabone} below. The computation times given in the table are in the units of seconds. The times are measured on a Dell Vostro 1700 laptop with dual cores of 1.8 GHz and 1GB RAM, which is used to run the MATLAB code of 50 realizations. As it can be seen on the table, for smaller number of time steps the CSSFM provides a small improvement in the computational effort. This is due to the computational effort required by the ${l_1}$ minimization. However for the bigger number of time steps, the computational effort is significantly reduced while the differences in the waveform are of negligible importance. Therefore the CSSFM provides a great computational efficiency compared to the classical SSFM and can be used as a tool in applied mathematics and physics.

\begin{table}[H]
\begin{center}
\caption{Comparison of Temporal Cost of the Classical SSFM vs Proposed CSSFM: One-Soliton Solution.\label{tabone}}
\vspace{10pt}
%\tbl{Comparison of Temporal Cost of the Classical Spectral vs Proposed Compressive Spectral Method.\label{tabone}}
{\begin{tabular}{@{}ccccccc@{}} \toprule
$N$ & $M$ & T. Steps & SSFM-T (s) & CSSFM-T (s) & Rms Diff.  \\
\hline
1024\hphantom{00} & \hphantom{0}64 & \hphantom{0} $10^5$& 308.75 & 73.04 &  0.0209 \\
1024\hphantom{00} & \hphantom{0}128 & \hphantom{0}$10^5$& 256.40 & 96.73 &  0.0025 \\
1024\hphantom{00} & \hphantom{0}128 & \hphantom{0}$10^6$& 2735.00 & 527.44 & 0.0025 \\
1024\hphantom{00} & \hphantom{0}256 & \hphantom{0}$10^5$& 302.81 & 182.43 & 0.0012 \\
2048\hphantom{00} & \hphantom{0}256 & \hphantom{0}$10^5$& 613.63 & 541.15 & 0.0023 \\
\hline
%0.1\hphantom{00} & \hphantom{0}876.0 & \hphantom{0}875.74 & 0.03 \\
%0.01\hphantom{0} & 2441.0 & 2441.0\hphantom{0} & 0.0\hphantom{0} \\
%0.001 & 4130.0 & 4129.3\hphantom{0} & 0.16\\ \botrule
\end{tabular} }
%\begin{tabnote}
%Table notes
%\end{tabnote}
%\begin{tabfootnote}
%\tabmark{a} Table footnote A\\
%\tabmark{b} Table footnote B
%\end{tabfootnote}
\end{center}
%\caption{Derivative relations for Hartley and Fourier transforms}
\end{table}
%\end{doublespace}

\subsection{Evolution of Two-Solitons}
%\begin{doublespace}
In the Figure~\ref{fig3} below, the $N=2048$ component SSFM and the $M=256$ component CSSFM are compared for two-soliton solution. The normalized root-mean-square difference between two profiles is $0.0032$ for this simulation. In the Figure~\ref{fig4} below, the  $N=1024$ component SSFM and the $M=256$ component CSSFM are compared for two-soliton solution.  The normalized root-mean-square difference between two profiles is $0.0068$ for this simulation. The two methods are in excellent agreement as it can be seen in the figures. The initial condition for these simulation is defined by (\ref{eq03})-(\ref{eq08}).

\begin{figure}[h]
\begin{center}
   \includegraphics[width=3.4in]{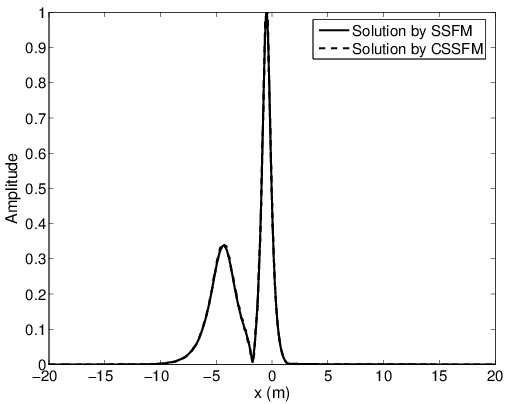}
  \end{center}
\caption{\small Comparison of SSFM with $N=2048$ and  CSSFM with $M=256$ components for two-soliton evolution.}
  \label{fig3}
\end{figure}

\begin{figure}[htb!]
\begin{center}
   \includegraphics[width=3.4in]{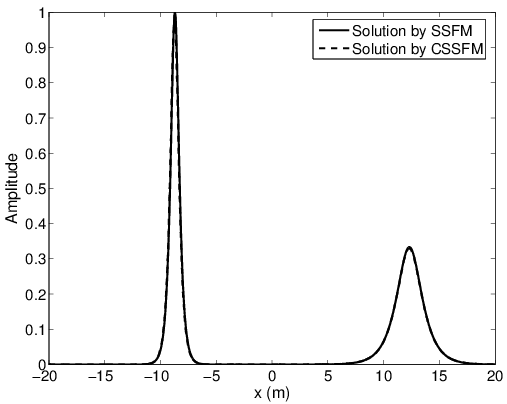}
  \end{center}
\caption{\small Comparison of SSFM with $N=2048$ and  CSSFM with $M=128$ components for two-soliton evolution.}
  \label{fig4}
\end{figure}

 All of the results presented in the figures above for two-soliton solution again show promising evidence for the accuracy of the proposed CSSFM. Additionally the computational effort required to run the various cases are summarized in the Table~\ref{tabtwo} below. The computation times given in the table are in the units of seconds. The times are measured on a Dell Vostro 1700 laptop with dual cores of 1.8 GHz and 1GB RAM which is used to run the MATLAB code of 50 realizations. Decreasing $M$, the spectral components of CSSFM, improves the efficiency however decreases the accuracy in an insignificant level. This is due to the reduced sampling points which affects the recovery of the sparse signal.  For smaller number of time steps the CSSFM provides only a small improvement in the computational effort. This is due to the computational effort required by the ${l_1}$ minimization. However for the bigger number of time steps such as $10^6$, the computational effort is significantly improved while the accuracy differences in the profiles are of negligible importance. Therefore the CSSFM provides a great computational efficiency compared to the classical SSFM and can be used as a tool in applied mathematics and physics.

\begin{table}[H]
\begin{center}
\caption{Comparison of Temporal Cost of the Classical SSFM vs Proposed CSSFM: Two Soliton-Solution.\label{tabtwo}}
\vspace{10pt}
%\tbl{Comparison of Temporal Cost of the Classical Spectral vs Proposed Compressive Spectral Method.\label{tabone}}
{\begin{tabular}{@{}ccccccc@{}} \toprule
$N$ & $M$ & T. Steps & SSFM-T. (s) & CSSFM-T. (s) & Rms Diff.  \\
\hline
1024\hphantom{00} & \hphantom{0}64 & \hphantom{0}$10^5$& 328.01 & 59.54 &  0.0160 \\
1024\hphantom{00} & \hphantom{0}128 & \hphantom{0}$10^5$& 257.04 & 107.00 &  0.0052 \\
1024\hphantom{00} & \hphantom{0}128 & \hphantom{0}$10^6$& 2607.40 & 501.90 & 0.0052 \\
1024\hphantom{00} & \hphantom{0}256 & \hphantom{0}$10^5$& 297.00 & 242.70 & 0.0034 \\
2048\hphantom{00} & \hphantom{0}256 & \hphantom{0}$10^5$& 614.39 & 532.88 & 0.0031 \\
\hline
%0.1\hphantom{00} & \hphantom{0}876.0 & \hphantom{0}875.74 & 0.03 \\
%0.01\hphantom{0} & 2441.0 & 2441.0\hphantom{0} & 0.0\hphantom{0} \\
%0.001 & 4130.0 & 4129.3\hphantom{0} & 0.16\\ \botrule
\end{tabular} }
%\begin{tabnote}

\end{center}
%\caption{Derivative relations for Hartley and Fourier transforms}
\end{table}
%\end{doublespace}

\section{Conclusion and Future Work}
In this study compressive split-step Fourier method (CSSFM) for the numerical simulation of the sparse signals is introduced . The sparsity property of the one- and two-soliton solutions of the nonlinear Schr\"{o}dinger equation is used for this purpose. 

It is shown that by using a smaller number of spectral components and the compressive sampling technique, it is possible to reconstruct time evolved signal with negligible difference compared to the classical split-step Fourier (SSFM) which uses a higher number of spectral components. It is shown that the proposed CSSFM improves the computational effort significantly. This improvement becomes more significant especially for large time evolutions. Therefore CSSFM can be used as an efficient tool in computational mathematics.

There are some sparse FFT algorithms developed in the literature. As a future work it is possible to implement these sparse fast transforms for computational modeling of the sparse signals and provide a comparison with the SSFM. It is also possible to perform an intercomparison of these models with the CSSFM proposed in this paper. The sequential, parallel or distributed algorithms can be used for this purpose.

The CSSFM can also be incorporated for other type of spectral methods such as those where the Legendre, Chebyshev and other forms of basis functions are used for computational simulations.


%\bibliographystyle{elsarticle}

%\bibliography{bibliog}

\begin{thebibliography}{99}

\bibitem{bay2009}  Bay\i nd\i r, C., (2009), Implementation of a Computational Model for Random Directional Seas and Underwater Acoustics, MS Thesis, University of Delaware.

\bibitem{bay_cssfm}
Bay\i nd\i r, C (2015). Compressive Split-Step Fourier Method. TWMS: Journal of Applied and Engineering Mathematics.  5, 298.


\bibitem{bay2015}  Bay\i nd\i r, C., (2015), Early detection of rogue waves by the wavelet transforms, Physics Letters A, 10.1016/j.physleta.2015.09.051.

\bibitem{bay_sun1} Bay\i nd\i r, C., (2015), Hesaplamal\i \ ak\i \c{s}kanlar mekani\u{g}i \c{c}al\i \c{s}malar\i \ i\c{c}in s\i k\i \c{s}t\i r\i labilir Fourier tayf\i \ y\"{o}ntemi, 19. Mekanik Kongresi, Trabzon (In Turkish).

\bibitem{bay_sun3}
Bay\i nd\i r, C (2015). S\"{o}n\"{u}ml\"{u} de\u{g}i\c{s}tirilmi\c{s} Korteweg de-Vries (KdV) denkleminin analitik ve hesaplamal\i \ \c{c}\"{o}z\"{u}m kar\c{s}\i la\c{s}t\i rmas\i.  19. Mekanik Kongresi. Trabzon, Turkey. (In Turkish)

\bibitem{bay_sun2} Bay\i nd\i r, C., (2015), Okyanus dalgalar\i n\i n s\i k\i \c{s}t\i r\i labilir Fourier tayf\i \ y\"{o}ntemiyle h\i zl\i \ modellenmesi, 19. Mekanik Kongresi, Trabzon (In Turkish).


\bibitem{bogomolov} Bogomolov, Y. L. and Yunakovsky, A. D., (2006), Split-step Fourier method for nonlinear {S}chrodinger equation, Proceedings of the International Conference Day on Diffraction, pp. 34-42.

\bibitem{candes} Candes, E. J., Romberg, J. and Tao, T., (2006), Robust uncertainty principles: Exact signal reconstruction from highly incomplete frequency information, IEEE Transactions on Information Theory, 52, pp. 489-509.

\bibitem{canuto} Canuto, C., Hussaini, M. Y., Quarteroni, A. and Zang, T. A. , (2006), Spectral Methods: Fundamentals in Single Domains, Springer-Verlag, Berlin.

\bibitem{demiray} Demiray, H. and Bayindir, C., (2015), A note on the cylindrical solitary waves in an electron-acoustic plasma with vortex electron distribution, Physics of Plasmas, 22, 092105; doi: 10.1063/1.4929863.

\bibitem{hardin} Hardin, R. H. and Tappert, F. D., (1973), Applications of the split-step Fourier method to the numerical solution of nonlinear and variable coefficient wave equation, SIAM Review Chronicles, 15, pp. 423-423.

\bibitem{hirota} Hirota, R., (1973), Exact envelope-soliton solutions of a nonlinear wave equation, The Journal of Mathematical Physics, 14, pp. 805-809.

\bibitem{Karjadi2010} Karjadi, E. A., Badiey, M. and Kirby, J. T., (2010), Impact of surface gravity waves on high-frequency acoustic propagation in shallow water, The Journal of the Acoustical Society of America, 127, pp. 1787-1787.

\bibitem{Karjadi2012} Karjadi, E. A., Badiey, M., Kirby, J. T. and Bayindir, C., (2012), The effects of surface gravity waves on high-frequency acoustic propagation in shallow water, IEEE Journal of Oceanic Engineering, 37, pp. 112-121.

\bibitem{taha} Taha, T. R. and Ablowitz, M. J., (1984), Analytical and numerical aspects of certain nonlinear evolution equations. II. Numerical Nonlinear {S}chrodinger Equation, Journal of Computational Physics, 22, pp. 203-230.

\bibitem{trefethen} Trefethen, L. N., (2000), Spectral Methods in {MATLAB}, SIAM, Philadelphia.

\bibitem{zakharov} Zakharov, V. E., (1968), Stability of periodic waves of finite amplitude on the surface of a deep fluid, Soviet Physics JETP, 2, pp. 190-194.


\end{thebibliography}

\end{document}